\documentstyle[prb,twocolumn,psfig,twoside,aps]{revtex}
\renewcommand{\vec}[1]{{\bf #1}}                                 
\renewcommand{\baselinestretch}{1.07}

\begin{document}
%
\addtolength{\oddsidemargin}{3ex}                                
\addtolength{\evensidemargin}{-2.5ex}                            
\addtolength{\textheight}{6ex}

\title{Ferromagnetism in the Hubbard model:\\ 
         Influence of the lattice structure}
\author{T. Herrmann and W. Nolting}
\address{Humboldt-Universit\"at zu Berlin, Institut f\"ur Physik, Invalidenstr.\ 110, 10115 Berlin, Germany}
\date{(15.Nov.1996, revised version 14.Apr.1997)}
\maketitle

\begin{abstract}
By use of the spectral density approach the influence 
of the lattice structure on the possibility of 
ferromagnetism in the single band Hubbard 
model is investigated. The $d=\infty$ hypercubic lattice 
does not show magnetic
phase transitions of second order irrespective of the 
strength of the
Coulomb coupling. However, first order transitions 
to finite magnetic moments,
not visible as singularities  of the paramagnetic 
susceptibility, may appear
in the very strong coupling regime.
In  d=3 second order  transitions are found but only 
for very strong couplings,
where the non-locality of the electronic self-energy 
acts in favour of the
spontaneous magnetic moment. The influence of the non-local part of
the self-energy is 
particularly strong for lattices with small
coordination number.  The non-bipartite fcc lattice exhibits
saturated ferromagnetism for all band occupations 
$1\le n\le 2$ while for less
than half filled bands ($0\le n\le 1$) the system remains in any case
paramagnetic, and that for $d=3$ as well as $d=\infty$. 
The Curie temperature
runs through a maximum at about $n=1.4$ and vanishes 
for $n\rightarrow 1$ and 
$n\rightarrow 2$. 
\end{abstract}

\pacs{PACS: 71.27.+a, 75.10Lp, 75.40.Cx}


\section{Introduction}\label{sec_introduction}
Originally the Hubbard model \cite{Hub63,Kan63,Gut63} was 
introduced to describe the bandmagnetism 
of itinerant electrons, at least qualitatively. 
The model Hamiltonian
\begin{equation} \label{eq_hamiltonian}
      {\mathcal H}=\sum_{i,j,\sigma}(T_{ij}-\mu)
      c_{i\sigma}^{\dagger}c_{j\sigma}
      +\frac{1}{2}U\sum_{i,\sigma}n_{i\sigma}n_{i-\sigma}
\end{equation}
incorporates, in the simplest way, the 
interplay of kinetic energy, Coulomb
interaction, band structure and Pauli principle 
for studying magnetic and
electronic properties of interacting electrons 
in a single non-degenerate narrow
energy band. $T_{ij}$ is the intersite hopping integral, 
usually taken between
nearest neighbors, while $U$ represents the Coulomb repulsion. 
$c_{i\sigma}\,(c_{i\sigma}^{\dagger})$ stands for the 
annihilation (creation)
operator of an electron with spin $\sigma=\uparrow,\,\downarrow$ 
at lattice
site $\vec{R}_{i}$. By $\mu$ we denote the chemical potential. 
$n_{i\sigma}=c_{i\sigma}^{\dagger}c_{i\sigma}$ is 
the occupation number
operator. The model-Hamiltonian (\ref{eq_hamiltonian}) 
defines a non-trivial
many-body problem that could be solved up to now 
only for some limiting cases
\cite{LW68,BHP73,Nag66,Lie89}. 

Although it was soon realized, that the Hubbard model is rather 
a generic model for antiferromagnetism, the question whether or not the
Hubbard model possesses ferromagnetic solutions 
continues to be the matter of controversial discussions.
In this work we focus
exclusively on ferromagnetic solutions of the single band Hubbard model. 
What concerns ferromagnetism
of real substances, as Fe, Co, Ni,  the direct Heisenberg 
exchange, Hund's rule coupling and band-degeneracy, of course, play
an important role \cite{Vol97}, the investigation of which is beyond the scope of
this paper.


The Mermin-Wagner theorem \cite{MW66} applied to 
the Hubbard model \cite{Gho71}
excludes ferromagnetism for one- and two-dimensional lattices. 
For the three
dimensional system with infinitely strong Coulomb repulsion
($U\rightarrow\infty$) exact results have been derived 
in a pioneering paper by
Nagaoka \cite{Nag66}. For very special band fillings 
$n_{\pm}=(N\pm1)/N$ ($N$:
number of lattice sites) ferromagnetic ground states 
are possible; in the case
of bipartite (sc and bcc) lattices for $n_{+}$ as well 
as for $n_{-}$, in the
non-bipartite lattice (fcc) only for $n_{+}$. 
This indicates that the
possibility of ferromagnetism crucially depends 
on the lattice structure. For
the hypercubic (hc) lattice at infinite dimensions 
Fazekas et. al. \cite{FMMH90}
showed that saturated ferromagnetism is excluded. 
However, in a recent study 
Ulmke \cite{Ulm96} applied a finite temperature 
Quantum Monte Carlo calculation
to the Hubbard model on a  fcc lattice at infinite 
spatial dimensions. A highly
polarized ferromagnetic ground state is found in a 
rather wide range of
band occupation. It is to be supposed that the greatest chance of
ferromagnetism to appear in the single band Hubbard 
model is for non-bipartite
lattices \cite{MHHH93,Uhr96}.

In this paper we want to investigate how sensible band 
ferromagnetism in the
Hubbard model depends on the lattice structure, 
and therewith on the density of
states of the non-interacting electrons 
(Bloch density of states: BDOS). We
apply a "spectral density approach" (SDA) the 
reliability of which, at least
for the strong coupling limit, has been demonstrated 
in previous papers
\cite{NB89,BdKNB90,HN96b}. The same approach has been 
applied to a somewhat
generalized model of magnetism (multiband Hubbard model) 
for the band ferromagnets 
Fe, Co, Ni \cite{NBDF89,NVF95,VN96}.
The SDA basically consists in a two-pole ansatz for 
the single-electron
spectral density, justified by the rigorous moment 
analysis of Harris and Lange
\cite{HL67}. The free parameter of the ansatz are 
self-consistently  fixed by
equating exactly calculated spectral moments. 
The SDA turns out to be
essentially equivalent to the Roth method \cite{Rot69,BE95} 
and to the Mori
projector formalism \cite{Mor65b,Zwa61}. 

For several types of BDOS we have derived the static 
paramagnetic susceptibility
$\chi$ as a function of the band occupation 
($0\le n\le 2$) and the temperature
$T$.  From the singularities we read off the 
instabilities with respect to
ferromagnetic order. The results are confirmed by a 
direct calculation of the
spontaneous magnetic moment. However, the 
$\chi^{-1}$-zeros are due to magnetic
phase transitions of second order only, while 
additional first order transitions to
a finite spontaneous moment may occur.

The SDA statements on the possibility of 
ferromagnetism in the Hubbard model
can be compared with  recent results from the $d=\infty$-technique
\cite{Ulm96,MHHH93,Uhr96}. As to the lattice 
structure dependence the SDA
solutions predict qualitatively the same trend. 
It is found, e.g., that for
the $d=\infty$-hypercubic lattice ferromagnetism 
is excluded,  being, on the
other hand, possible for an $d=\infty$-fcc lattice.

For finite spatial dimensions ($d=3$)
we find that the non-locality of the quasiparticle self-energy, 
vanishing for
$d\rightarrow\infty$ in the SDA, too, has a stabilizing 
influence on the
magnetic order. As an example, the  $d=\infty$ hypercubic lattice is
non-magnetic, but the $d=3$ sc lattice becomes ferromagnetic for
$U/W\ge U_{c}/W =4$ and $0.34\le n\le 1.66$ if the full 
$\vec{k}$-dependent self-energy is taken into account. Neglecting the 
$\vec{k}$-dependent part of the
self-energy drives $U_{c}/W$ to a higher value 
($=14$) with $n_{c}$ only
slightly changed.

\section{Spectral density approach}\label{sec_sda}
The method is based on a physically  motivated ansatz 
for the single-electron 
spectral density, which is defined by:
\begin{eqnarray}
	S_{ij\sigma}(E)&=&\int\limits_{-\infty}^{+\infty}\!\!dt
	  e^{-\frac{i}{\hbar}Et} S_{ij\sigma}(t)
	  \label{eq_spectral_density_E}\\
	S_{ij\sigma}(t)&=& \frac{1}{2\pi}\langle
        [c_{i\sigma}(t),c_{j\sigma}^{\dagger}(0)]_{+}\rangle
        \label{eq_spectral_density}
\end{eqnarray}
$[..\,,..]_{+}$ is the anticommutator, and 
$\langle\dots\rangle$ means thermodynamic
averaging. The operators in (\ref{eq_spectral_density}) 
are thought as
time-dependent Heisenberg operators. The mentioned ansatz 
contains some free
parameters which can be fitted by equating a set of spectral moments
$M_{ij\sigma}^{(n)}$
\begin{equation}
	M_{ij\sigma}^{(n)}=\int\limits_{-\infty}^{+\infty}\!\!dE
	E^{n} S_{ij\sigma}(E);\qquad n=0,1,2,\dots
	\label{eq_moments}
\end{equation}
The moments are calculated exactly via 
\begin{equation}
	M_{ij\sigma}^{(n)}=
	    \langle\bigg[\underbrace{\Big[
	    \dots[c_{i\sigma},{\mathcal H}]_{-},\dots,
	    {\mathcal H}\Big]_{-}}_{\textrm{$n$-fold commutator}}
          ,c_{j\sigma}^{\dagger}\bigg]_{+}\rangle
          \label{eq_moments2}		
\end{equation}	
In ref. \cite{HN96b} a two-pole ansatz for the spectral 
density of the strongly
coupled Hubbard model is justified as a reasonable 
starting point requiring the
fitting of the first four spectral moments.  
The evaluation leads to an electronic self-energy
of the following structure:
\begin{equation}
	\label{eq_self-energy}
	\Sigma_{\vec{k}\sigma}^{\textrm{\scriptsize{SDA}}}(E)=
	U\langle n_{-\sigma}\rangle
	\frac{E+\mu-B_{-\sigma}-F_{\vec{k}-\sigma}}
	{E+\mu-B_{-\sigma}-F_{\vec{k}-\sigma}-
	 U\left(1-\langle n_{-\sigma}\rangle \right)}.
\end{equation}
The decisive terms are $B_{-\sigma}$ and $F_{\vec{k}-\sigma}$.
For $B_{-\sigma}=F_{\vec{k}-\sigma}=0$  (\ref{eq_self-energy}) 
reproduces the Hubbard-I
solution \cite{Hub63}, which is incapable to describe ferromagnetism. 
$B_{-\sigma}$ as well as $F_{\vec{k}-\sigma}$ consists 
of "higher" correlation
functions, possibly  creating a spin-dependent shift 
of the Hubbard bands and
therewith resulting in a finite spontaneous magnetic moment.
The  term $B_{-\sigma}$, 
\begin{equation}   
	\label{eq_bsigma}
	\langle n_{-\sigma}\rangle (1-\langle n_{-\sigma}\rangle )
	B_{-\sigma}=\frac{1}{N}
	\sum_{i,j}^{i\neq j} T_{ij}\langle c_{i-\sigma}^{\dagger} 
	c_{j-\sigma}(n_{i\sigma}+n_{j\sigma}-1)\rangle,
\end{equation}
can rigorously be expressed by the single-electron spectral density
\cite{NB89,HN96b}. The other term,
\begin{eqnarray}   
	\label{eq_fsigma}
	\langle n_{-\sigma}\rangle (1-\langle n_{-\sigma}\rangle )
	F_{\vec{k}-\sigma}&=& \nonumber\\
	& &\hspace{-2.5cm}\frac{1}{N}\sum_{i,j}^{i\neq j} T_{ij}
	e^{-i\vec{k}\cdot(\vec{R}_{i}-\vec{R}_{j})}
	\Big[(\langle n_{i-\sigma} 
	n_{j-\sigma}\rangle -n_{-\sigma}^{2})
	                        \nonumber\\
	& &\hspace{-2.5cm} -\langle c_{j\sigma}^{\dagger} c_{j-\sigma}^{\dagger} 
	     c_{i-\sigma} c_{i\sigma}\rangle 
	- \langle c_{j\sigma}^{\dagger} c_{i-\sigma}^{\dagger}
	  c_{j-\sigma} c_{i\sigma}\rangle\Big],
\end{eqnarray}
incorporates a density-density, a double hopping-, 
and a spinflip-correlation.
For the evaluation of these expectation values properly 
modified spectral densities,
\begin{equation}
	S_{ij\sigma}^{(A)}(t)=\frac{1}{2\pi}
	  \langle [A_{i}(t),c_{j\sigma}^{\dagger}(0)]_{+}\rangle,
	\label{eq_spectral_density_A}
\end{equation}
are introduced, where the operator $A_{i}$ is chosen 
in such a way that the
respective expectation value can be derived directly 
via the spectral theorem
from $S_{ij\sigma}^{(A)}(E)$. It  can be justified by 
inspection of the Lehmann
representation  of these functions that in the wave-vector 
representation
they  must represent  two-pole
functions as the original single-electron spectral density 
(\ref{eq_spectral_density_E}). The various functions 
(\ref{eq_spectral_density_E}) and 
(\ref{eq_spectral_density_A})  differ  only
by the spectral weights of the poles. For fixing 
$S_{ij\sigma}^{(A)}(E)$
one therefore needs only two respective moments. 
Doing so one finally
arrives at a closed system of equations that can be 
solved self-consistently for
all quantities of interest. For further details the 
reader is referred to
\cite{HN96b}.

We study the magnetic phase transition by use of the 
static susceptibility
\begin{equation}
	\label{eq_susceptibility}
	\chi(T,H,n)=\frac{1}{\mu_{0}}
	\left(\frac{\partial M}{\partial H}\right)_{T,n} .
\end{equation}
$\mu_{0}$ is the vacuum permeability and $H$ a 
homogeneous static magnetic
field. $M$ denotes the magnetization
\begin{equation}
	M=\frac{N}{V}\mu_{B}
	(\langle n_{\uparrow} \rangle -
	 \langle n_{\downarrow} \rangle ) .
\end{equation}
We calculate $\chi$ for the paramagnetic system in the 
zero-field limit. The
poles of the susceptibility indicate the instabilities 
of the system towards
ferromagnetic ordering. To calculate the static susceptibility 
$\chi$ we have
to include a Zeeman term in the Hamiltonian (\ref{eq_hamiltonian}).
When differentiating $M$ with respect to the external 
field one has to bear in
mind that all expectation values 
$\langle\dots\rangle$
are field-dependent.

\section{Results}\label{sec_results}
\begin{figure}[htb]
	\centerline{\psfig{figure=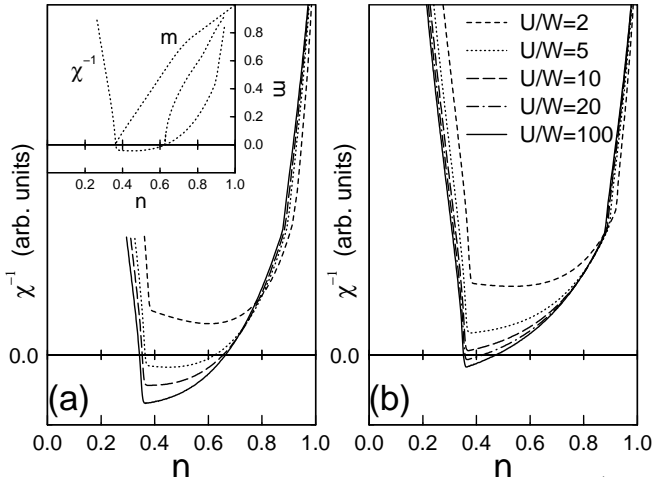,height=7cm,angle=270}}
	\caption{Inverse paramagnetic static susceptibility 
	$\chi^{-1}$ for 
	the sc lattice ($d=3$) as a function of
	the band occupation $n$ for various values of the 
	Coulomb interaction $U$. 
	(a) System with the full $\vec{k}$-dependent self-energy. 
	(b) System with a 
	local self-energy ($F_{\vec{k}-\sigma}\equiv0$ artificially, 
	see eq.~(\ref{eq_self-energy})).
	The inset in (a) shows the magnetization curves $m(n)$ 
	of the two ferromagnetic 
	solutions starting at the two zeros of  
	$\chi^{-1}$ for $U/W=5$. 
	($T=0\,$K).
\label{fig_one}} 
\end{figure}

We have investigated the tendency to ferromagnetism 
in the single band Hubbard 
model for several lattice structures and different 
spatial dimensions. The
zeros of the  inverse static paramagnetic 
susceptibility indicate the magnetic
phase transitions of second order. Fig.~\ref{fig_one} 
shows the results for a
sc lattice. 
Sometimes it is argued \cite{SV95} that 
for this bipartite lattice
ferromagnetism is hardly to be expected. On the other hand, 
Shastry et.~al.~\cite{SKA90} report for the $d=3$ model even 
saturated
ferromagnetism for $n\ge 0.68$. That is very close to our 
findings. It is a
special feature of the SDA \cite{NB89,BdKNB90,HN96b}, 
possibly even for the
Hubbard model itself, that there are two ferromagnetic 
solutions, i. e. two
zeros of $\chi^{-1}$. The first solution sets in at 
$n_{c}^{(1)}=0.34$
($U\rightarrow\infty$), where the actual value only 
slightly depends on $U$.
This solution runs into saturation for $n\ge0.68$ 
($U\rightarrow\infty$), in
exact agreement with the results of ref.~\cite{SKA90}. 
The second solution
appears for higher band occupations, but does never 
reach saturation. This less
magnetized solution is always less stable and can be 
disregarded in the
following discussion. In spite of  the appearance of 
ferromagnetic solutions
the sc structure seems not to be very convenient for 
a spontaneous order. A
rather strong Coulomb coupling $U/W>4$ (W: Bloch bandwidth) 
is needed. In this
context it is interesting to look at the influence 
of the non-locality of the
electronic self-energy (\ref{eq_self-energy}).  
The wave-vector  dependence 
comes into play due to the higher correlation 
functions in $F_{\vec{k}-\sigma}$
(\ref{eq_fsigma}). Since in the limit of infinite 
spatial dimensions the
electronic self-energy is wave-vector independent \cite{MV89}, 
the role of
the $\vec{k}$-dependent terms in 
$\Sigma_{\vec{k}\sigma}^{\textrm{\scriptsize{SDA}}}(E)$ 
might be underestimated
for $d=3$. In part (b)  of Fig.~\ref{fig_one}  
we have plotted $\chi^{-1}$  as
function of the electron density  for the case 
that the self-energy
(\ref{eq_self-energy})  has been made local simply by setting 
$F_{\vec{k}-\sigma}\equiv 0$.  The critical couplings $U_{c}/W$ 
 for ferromagnetic order
rises enormously  from $4$ for the full 
$\vec{k}$-dependent self-energy 
(Fig.~\ref{fig_one}(a)) to $14$ for the local 
self-energy (Fig.~\ref{fig_one}(b)). 
Its influence on magnetic stability, however,  
is not for all cubic lattices of
the same importance. It can be recognized that with 
increasing number of
nearest neighbours the importance of the 
$\vec{k}$-dependent  part of the
self-energy  rapidly decreases.
For the  fcc lattice ($Z_{1}=12$)  it appears 
already  non-remarkably
\cite{BdKNB90}. 

\begin{figure}[htbp]
	\centerline{\psfig{figure=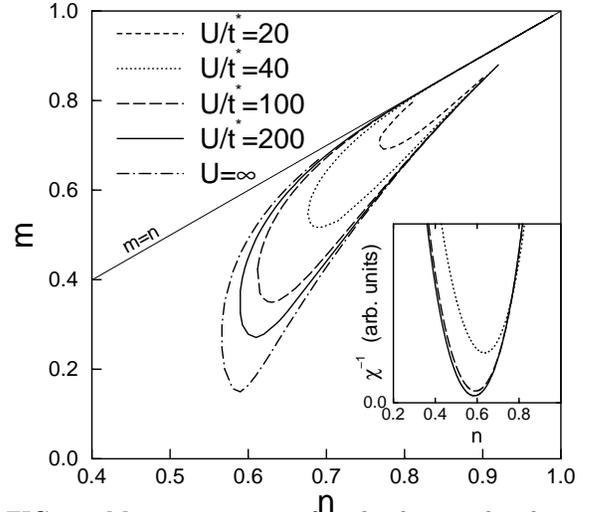,height=7cm,angle=270}}
 	\caption{Magnetization $m$ for the hypercubic 
 	lattice in $d=\infty$  
	as a function of the band occupation $n$ for various values 
	of the Coulomb interaction $U$. The BDOS is given by: 
	$\rho_{0}(E)=\exp(-0.5(E/t^{*})^{2})/(t^{*}\protect\sqrt{2\pi})$.
	Inset: Inverse para\-magnetic static 
	susceptibility $\chi^{-1}$ as a function of 
	the band occupation $n$.
	($T=0\,$K). \label{fig_two}}
\end{figure}

In infinite spatial dimensions, for which the higher correlation 
$F_{\vec{k}-\sigma}$ disappears, we do not find zeros 
in $\chi^{-1}$ for the
hypercubic lattice (inset of Fig.~\ref{fig_two}), and 
that irrespective of the
strength of the Coulomb coupling $U$.  Phase transitions 
of second order are therefore excluded. 
The same fact has been reported by 
Jarrell \cite{Jar92} who
elaborated a self-consistent Monte Carlo procedure, 
which is claimed to be
essentially exact for the $d=\infty$  Hubbard model. 
Although there is no
indication for ferromagnetism in the static susceptibility, 
the direct
evaluation of the SDA for the spin-dependent, 
averaged particle numbers
$\langle n_{\uparrow}\rangle$, $\langle n_{\downarrow}\rangle$  
reveals
ferromagnetic solutions (Fig.~\ref{fig_two}).  
They belong to first order
transitions being therefore not visible  as $\chi^{-1}$-zeros. 
These solutions appear for
rather strong Coulomb couplings and for band occupations $n\ge 0.57$. For a
given $U$ the $m(n)$ curve forms a closed bubble steadily 
contracting itself
for increasing temperature.  For a given band occupation 
$n$ this means a first
order transition at a certain critical temperature. 
Ferromagnetic saturation is, strictly speaking,
never reached, not even for $T=0K$ (Fig.~\ref{fig_two}). This is
because of the infinite tails of the Gaussian BDOS.
The strict physical meaning of this kind of ferromagnetism 
in the Hubbard
model requires further investigation.
 
\begin{figure}[htbp]
	\centerline{\psfig{figure=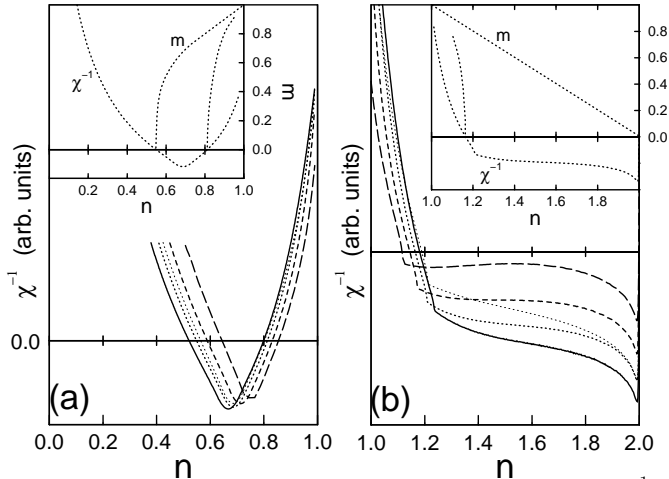,height=7cm,angle=270}}
	\caption{(a) Inverse paramagnetic static susceptibility 
	$\chi^{-1}$  
	for the bcc lattice as a function of
	the band occupation $n$ for various values of the 
	Coulomb interaction $U$ 
	(parameters as in Fig.~\ref{fig_one}(b)). 
	The thin dotted line corresponds to the system with  
	the local self-energy ($F_{\vec{k}-\sigma}\equiv0$ artificially, 
	see eq.~(\ref{eq_self-energy})) for $U/W=5$. 
	The inset  shows the magnetization curves $m(n)$ together with
	$\chi^{-1}$ for $U/W=5$.
	(b) the same as in (a) for the fcc lattice.
	($T=0\,$K).
\label{fig_three}}
\end{figure}

It turns out that the three dimensional bcc lattice 
with its divergence at the
center of the  BDOS is much more convenient  for 
ferromagnetism  than the sc
lattice. Already for $U/W\approx1$ two ferromagnetic 
solutions appear, the
stable one starts for $U\rightarrow\infty$ at $n_{c}=0.52$ 
reaching saturation
(m=n) for $n\ge 0.68$ (Fig.~\ref{fig_three}). 
The latter  result is again in
exact agreement  with the findings in ref.~\cite{SKA90}, 
where the stability of
the Nagaoka state \cite{Nag66} ($U/W\rightarrow\infty$) 
with respect to an
electron spin flip is investigated. As  already mentioned the
$\vec{k}$-dependence of the electronic self-energy has not 
such dramatic
consequences as for the sc lattice (Fig.~\ref{fig_one}).

The non-bipartite fcc lattice does not allow 
ferromagnetism for  less than
half-filled energy bands. However, for the more than 
half-filled  band we find
saturated ferromagnetism ($m=2-n$: "Nagaoka state") 
for all electron densities
$1.0\le n\le2.0$ indicated by a zero of $\chi^{-1}$ 
at $n=2.0$ (Fig.~\ref{fig_three}).
This agrees with the investigation of 
M\"{u}ller-Hartmann et.~al.~\cite{MHHH93}
who employed the single spin flip Gutzwiller variational 
states of Shastry
et.~al.~\cite{SKA90} and corrected their value of 
$n_{c}=1.62$ in \cite{SKA90}
to $n_{c}=2.0$. For electron densities 
$1.0\le n\le 1.2$ the SDA yields a
second ferromagnetic solution which is less 
stable than the fully polarized
one.
\begin{figure}[htbp]
	\centerline{\psfig{figure=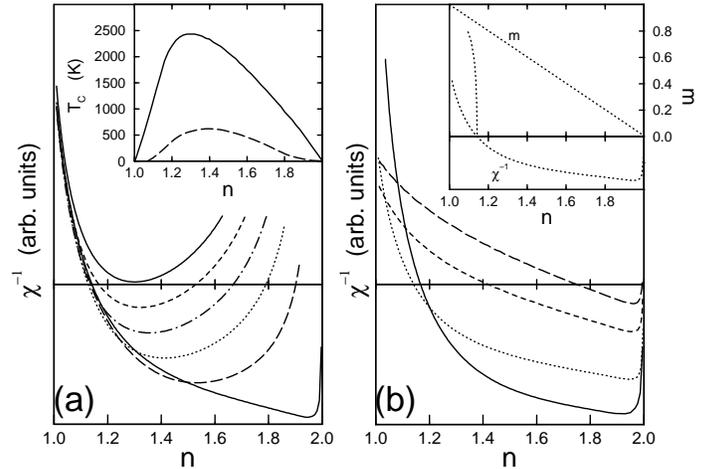,height=7cm,angle=270}}
	\caption{fcc lattice in $d=\infty$ 
	(BDOS: $\rho_{0}(E)=\exp(-0.5(1+\protect\sqrt{2}E))/
	\protect\sqrt{\pi(1+\protect\sqrt{2}E)}$ 
	from ref.~\protect\cite{Ulm96})
	(a) Inverse paramagnetic static susceptibility $\chi^{-1}(n)$  
	as a function of the band occupation $n$ for various 
	temperatures $T$ (solid: $T=0\,$K, long dashed: $T=500\,$K, 
	dotted: $T=1000\,$K, dot dashed: $T=1500\,$K, 
	dashed: $T=2000\,$K,
	dot dot dashed: $T=2500\,$K.)
	The inset shows the Curie temperature $T_{C}$ as a 
	function of the $n$. The solid line is the result of the SDA, the 
	long dashed line is taken from ref.~\protect\cite{Ulm96}. 
	($U=4\,$eV).
	(b) $\chi^{-1}(n)$  
	for various values of the 
	Coulomb interaction $U$. 
	The inset  shows the magnetization curves $m(n)$ 
	together with $\chi^{-1}$ 
	for $U=4\,$eV. (long dashed: $U=0.4\,$eV, dashed: $U=1\,$eV,
	dotted: $U=4\,$eV, solid: $U=100\,$eV; $T=0\,$K).
	\label{fig_four}}
\end{figure}

The results are qualitatively similar for the 
Hubbard model on the infinite
dimensional fcc lattice (Fig.~\ref{fig_four}). 
Here, we used the BDOS given in
ref.~\cite{Ulm96}. 
The $\chi^{-1}(n)$ curves 
exhibit two zeros due to two
ferromagnetic solutions as in the $d=3$ case 
(Fig.~\ref{fig_three}). The
fully polarized  state ($m=2-n$) 
is stable for all band
occupations $1.0\le n\le2.0$ with a critical coupling 
$U_{c}(n=2.0)/W=0^{+}$.
The same is reported in ref.~\cite{Ulm96,Uhr96}. 
The asymmetric BDOS of the
non-bipartite fcc lattice is obviously very convenient for the band
ferromagnetism in the more than half filled 
single band Hubbard model.

For finite temperature the $\chi^{-1}$-zero 
shifts away from $n=2.0$ to lower
values indicating a Curie-temperature of $T_{C}=0^{+}$ for $n=2$ 
(Fig.~\ref{fig_four}(a)). The $T_{C}(n)$ 
curve runs through a maximum at
$n\approx 1.4$ and goes to zero  for the half-filled 
($n\rightarrow1.0$) and
the fully occupied ($n\rightarrow2.0$) energy band 
(see inset in 
Fig.~\ref{fig_four}(a)). These results are 
qualitatively consistent with that
of Ulmke~\cite{Ulm96}. 
However, contrary to ref.~\cite{Ulm96}, our  
$T_{C}(n)$ curve persists until $n=1$. This reflects the fact, that
close to half filling ($1.0<n<1.15$) the  magnetization 
curves in the SDA exhibit first order phase transitions 
as a function of temperature. Magnetic solution of this kind  
where not considered  in ref.~\cite{Ulm96}.
Quantitatively our $T_{C}$-values are on an average
higher by a factor $4$, but nevertheless in 
a reasonable order of magnitude.

\section{Conclusions}\label{sec_conclusions}
By use of a spectral density approach we have 
demonstrated the striking
influence of the lattice structure on the possibility  
of ferromagnetism in the
Hubbard model. The zeros of the inverse static
 paramagnetic susceptibility
indicate the onset of ferromagnetic order in 
dependence of typical variables as
the band occupation $n$ and temperature $T$. 
In the hc lattice no "normal"
ferromagnetism appears at infinite dimensions, 
while for $d=3$ (sc lattice) magnetic
solutions are found if $U/W$ exceeds a rather 
high critical value. For  small lattice coordination number (e.g. sc lattice)
the non-locality of the electronic self-energy favours 
ferromagnetism in a remarkable manner. In the non-bipartite 
fcc lattice no magnetism exists for less than 
half-filled bands, while for  $1.0\le n\le2.0$ saturated 
ferromagnetism is found at $d=3$ as well as $d=\infty$.

\acknowledgments{
This work has been done within the Sonderforschungsbereich 290 ("Metallische
d\"{u}nne Filme: Struktur, Magnetismus und elektronische Eigenschaften") of the
Deutsche Forschungsgemeinschaft.}
\renewcommand{\baselinestretch}{1.01}

\end{document}